\documentclass[prb,superscriptaddress,twocolumn]{revtex4}
\usepackage{graphicx}
\usepackage{amsmath}
\usepackage{amssymb}
\usepackage{epsfig}
\usepackage{wasysym}
\usepackage{bbm}
\usepackage{multirow}
\renewcommand{\narrowtext}{\begin{multicols}{2} \global\columnwidth20.5pc}
\renewcommand{\v}[1]{{\bf #1}}

\newcommand{\s}{{\sigma}}

\newcommand{\gr}{{\nabla}}
\def\be{\begin{eqnarray}}
\def\ee{\end{eqnarray}}

\newcommand{\Eq}[1]{Eq.~(\ref{#1})}

\newcommand{\ra}{\rightarrow}

\newcommand{\e}{\epsilon}
\newcommand{\Fig}[1]{Fig.~\ref{#1}}

\newcommand \ti[1]{}

\begin{document}

\title{The Topological Relation Between Bulk Gap Nodes and Surface Bound States : Application to  Iron-based Superconductors}

\author{Fa Wang}
\affiliation{Department of Physics, Massachusetts Institute of Technology, Cambridge, Massachusetts 02139, USA}

\author{Dung-Hai Lee}
\affiliation{Department of Physics, University of California at Berkeley, Berkeley, CA 94720, USA}
\affiliation{Materials Sciences Division, Lawrence Berkeley National Laboratory, Berkeley, CA 94720, USA}



\begin{abstract}

In the past few years materials with protected gapless surface (edge) states have risen to the central stage of condensed matter physics. Almost all discussions centered around topological insulators and superconductors, which possess full quasiparticle gaps in the bulk. In this paper we argue systems with {\em topological stable} bulk nodes offer another class of materials with robust gapless surface states. Moreover the location of the bulk nodes determines the Miller index of the surfaces that show (or not show) such states. Measuring the spectroscopic signature of these zero modes allows a phase-sensitive determination of the nodal structures of unconventional superconductors when other phase-sensitive techniques are not applicable. We apply this idea to gapless iron-based superconductors and show how to distinguish accidental from symmetry dictated nodes. We shall argue the same idea leads to a method for detecting a class of the elusive spin liquids.
\end{abstract}
\maketitle

The interest in topological insulators and superconductors lie mainly in studying their robust gapless surface states\cite{review1,review2}. Systems with bulk gap nodes are usually neglected for it is generally felt bulk nodes render surface states ill defined. Gap nodes in the form of band crossing are singularities in the electronic structure. Examples include the Dirac nodes in graphene, the d-wave superconducting gap nodes of cuprates and the Weyl nodes\cite{mur,vish}. These singularities have dramatic effects on the boundary electronic structure. For example, the zigzag edge of graphene has zero energy (E=0) flat bands\cite{Lee,nakada}, and the \{110\} surfaces of the cuprate superconductors have zero bias Andreev bound states (ZBABS)\cite{hu,Tanaka-cuprate}. Recently it has been shown that the Weyl nodes will give rise to surface ``Fermi arcs''\cite{vish}. We shall argue the above examples are special cases of a more general fact, namely translational invariant systems with {\em topologically stable} bulk nodes also possess protected  gapless surface bound states. In the presence of disorder the otherwise degenerate zero energy band broadens into finite width, but does not lose the signature of gapless surface bound state.

An important class of ``nodal material'' is nodal superconductor. Usually, the presence of gap nodes reflects the overall repulsive nature of the (effective) Cooper pairing Hamiltonian. The location of the gap nodes provide detailed information about the momentum dependence of the pairing interaction. This information is necessary for the understanding of the pairing mechanism of nodal, hence unconventional, superconductors. Among these superconductors the pairing symmetry of iron based superconductors (FeSCs) is still very much under debate. Experimentally roughly half of the materials in this family show the evidence of gap nodes. Examples include LaFePO\cite{LaFePO}, BaFe$_2$(As$_{1-x}$P$_{x}$)$_2$\cite{BFAP1,BFAP2,BFAP3,feng}, KFe$_2$As$_2$\cite{KFe2As2a,KFe2As2b}, LiFeP\cite{LiFeP} etc. The nodal structures of the majority in this group are unknown. The notable exception is BaFe$_2$(As$_{1-x}$P$_{x}$)$_2$ whose node locations have been determined by angle-resolved photoemission spectroscopy\cite{feng}. Because of the complex fermiology and the likelihood of ``accidental nodes'' these materials are not accessible to conventional phase-sensitive probes. Finding experimental means to pin down the location of the nodes for these materials has been a major experimental challenge. In the following we propose to study the correlation between the orientation of the surfaces with the presence/absence of zero energy peaks in surface sensitive spectroscopy. We show this method does not suffer the above difficulty. We will illustrate the idea by focusing on LaFePO.  Near the end we shall briefly discuss gapless non-centrosymmetric superconductors, and generalize the idea presented here for the detection of a class of spin liquids.

{\bf{Bulk nodes and zero energy surface bound states.}}
The gap nodes of a d-dimensional quasiparticle Hamiltonian can have non-zero dimension $q$. For examples  $q=0$ for point node and $q=1$ for line node etc. A prerequisite of a stable nodal structure is in $d-q-1$ dimension the symmetry of the Hamiltonian\cite{ryu,kitaev} {\em at a generic $\v k$ point} admits topological insulator/superconductors\cite{sato,beri}. {\em If a nodal structure is stable},
and if $d-q-1>0$,
then suitably oriented boundaries of this material will have E=0 bound state. Generically the dimension of the sub boundary BZ possessing E=0 bound states is $d_{E=0}=q+1$. For graphene $d=2$, $q=0$; for cuprates $d=3$, $q=1$; and for the Weyl semimetal\cite{vish} $d=3$, $q=0$.
For the purpose of this paper we focus on the cases where the band topology of the $d-q-1$ dimensional system is classified by the group of integers\cite{ryu}. Under this condition a winding number\cite{ryu}(or vorticity) can be assigned to each independent nodal manifold ({e.g.} point or line). {\em As long as the projection of opposite-winding-number nodal manifolds does not completely overlap in the boundary Brillouin zone (BZ), there will be gapless surface bound states.}
Special cases of the above argument have already been put forward in Ref.~\cite{vish,Sato-flatband,ryu2}, 
especially in Volovik {et al.}\cite{Volovik}.
A more illustrative example is given in Appendix~\ref{app:example}.\\

\begin{figure}[tbp]
\begin{center}
\includegraphics[scale=0.44]{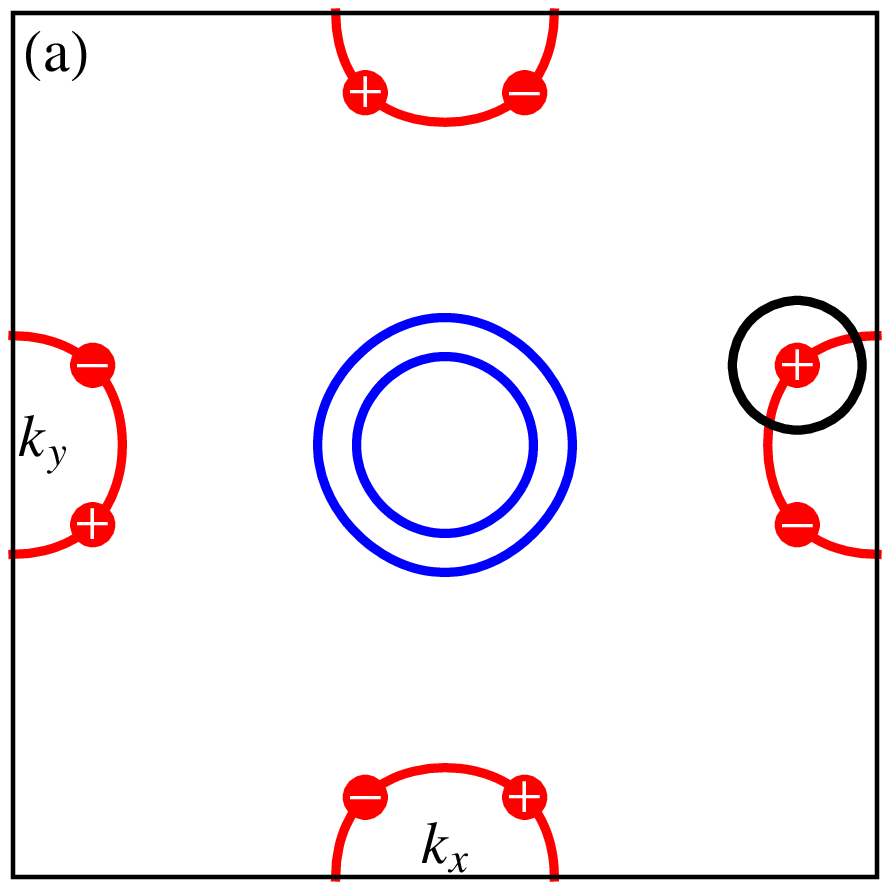}\hspace{0.01in}
\includegraphics[scale=0.44]{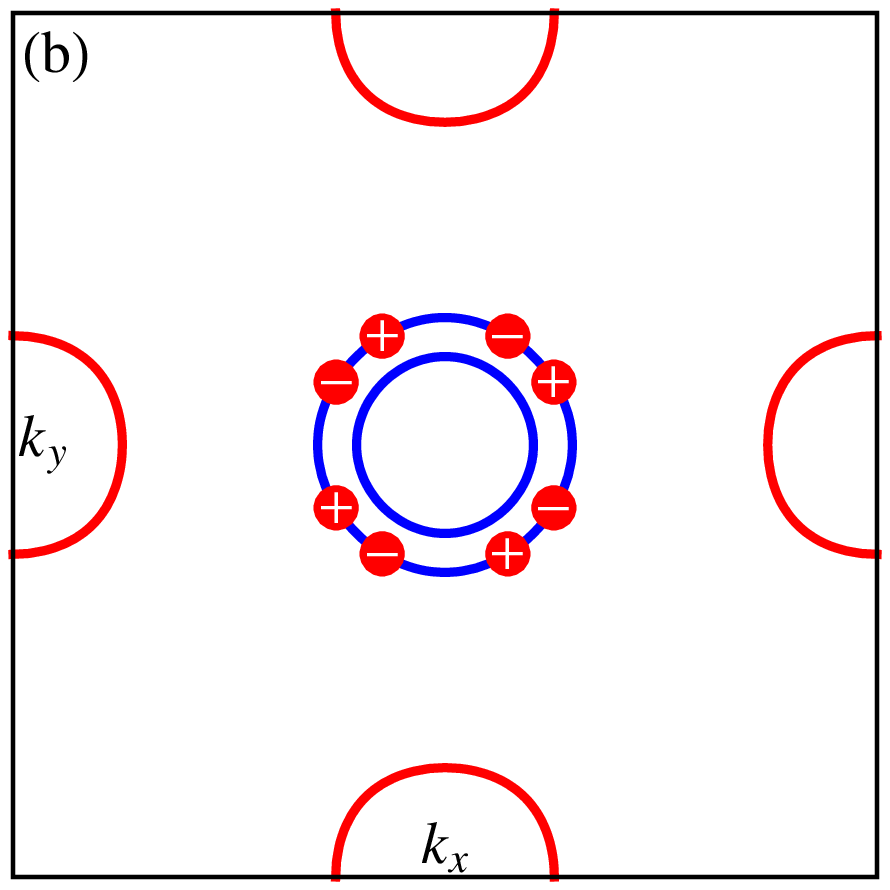}\\
\includegraphics[scale=0.44]{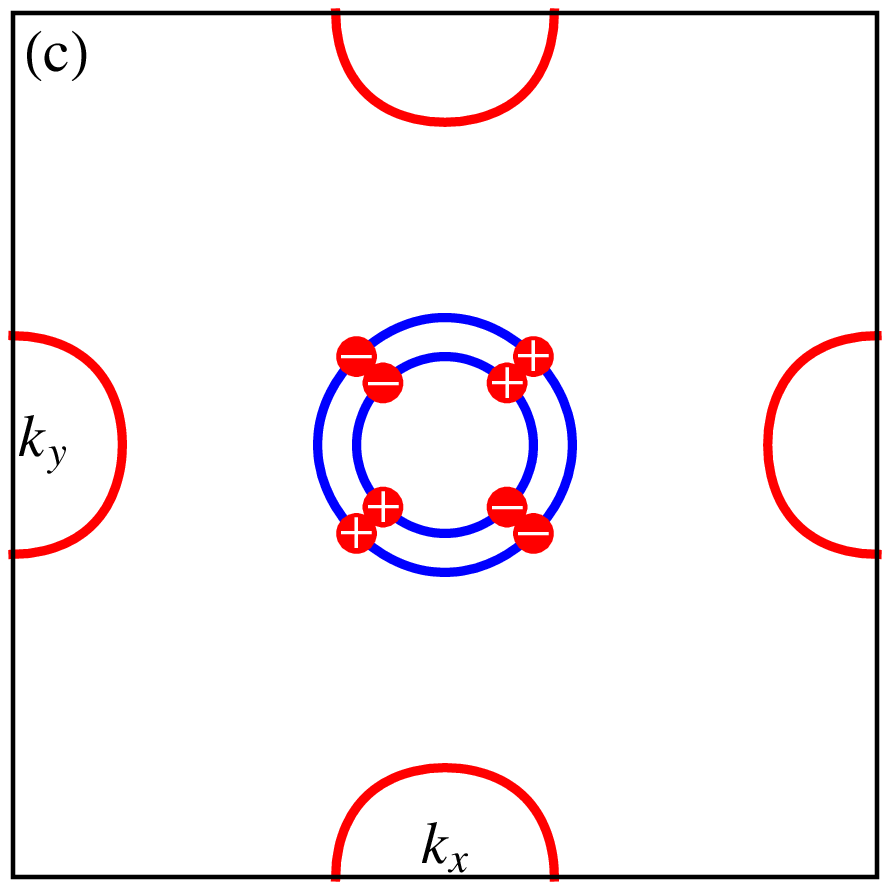}\hspace{0.01in}
\includegraphics[scale=0.44]{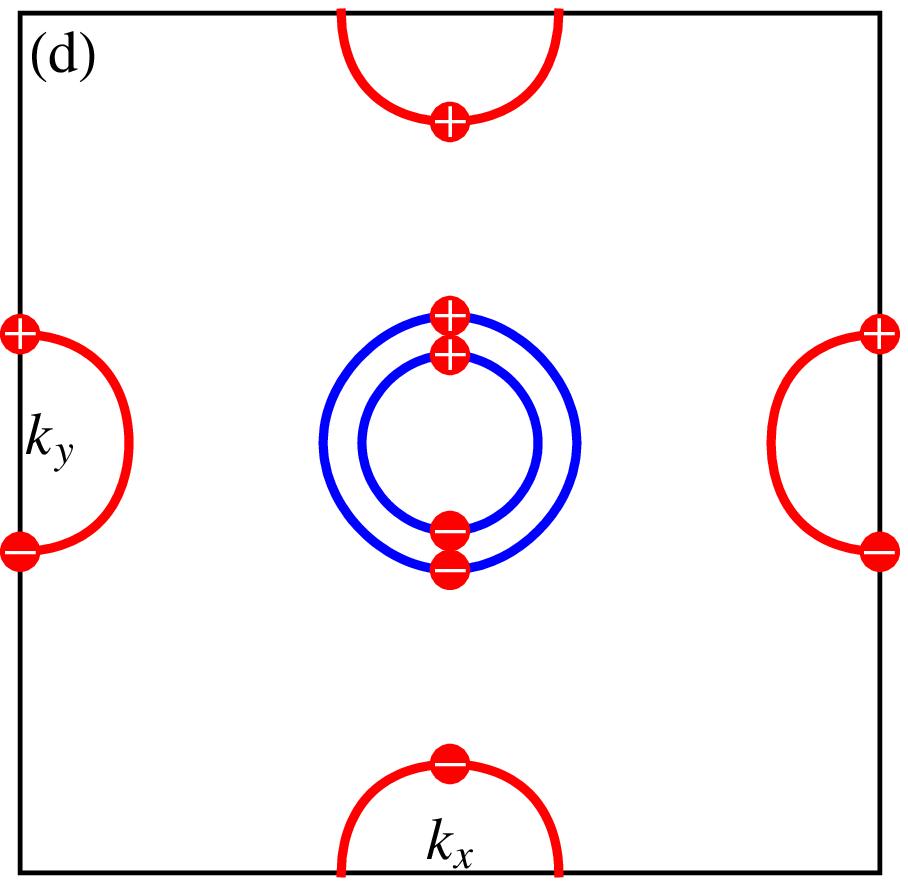}
\caption{Caricatures of the Fermi surfaces of LaFePO and four examples of nodal structures. The blue/red curves represent the hole/electron Fermi surfaces.
Red dots with $\pm$ signs indicate nodes with $\pm$ sign of winding numbers. 
(a,b) Two different $s$-wave gap nodes (red dots),
(c) $d_{x^2-y^2}$ nodes and (d) $p_x$ nodes.
The black circle in part (a) is a close loop enclosing a gap nodes.}
\label{fig1}
\end{center}
\end{figure}

{\bf{Application to LaFePO.}}
LaFePO is a stoichiometric compound, it is clean enough for quantum oscillation to be observed\cite{analytis}. According to Ref.~\cite{analytis}  LaFePO has nearly two dimensional Fermi surfaces: two hole pockets at the center and two electron pockets on the face of the unfolded BZ (for a caricature see \Fig{fig1}a). In the following we shall used 2D notations. LaFePO is the first FeSC where strong evidence of line nodes have been reported\cite{LaFePO}. At the present time there is no direct information about the location of the line nodes. In addition, early Knight shift measurement has revealed a rather puzzling result. Unlike conventional singlet-paired superconductors, the Knight shift increases rather than decreases below $T_c$\cite{nakai,comm}. This unusual behavior calls for the examination of possible nodal triplet pairing.

Theoretically it has been proposed that the {\em singlet} gap function of LaFePO possesses the so-called ``accidental nodes''\cite{acc1,acc2,acc3}  where the nodal gap function transforms without any sign change under all point group operations.
A schematic illustration of the proposed nodal structure is shown in \Fig{fig1}a.
First let's discuss the stability of these accidental nodes. Consider a loop encircling one of nodes (e.g. the black circle in \Fig{fig1}a). Like the cuprates the BdG Hamiltonian defined on the loop is characterized by an even integer winding number, and has a fully gapped spectrum. Thus the loop has a non-trivial band topology.
If the node is removed from the interior of the circle the winding number must change, which requires the spectrum on the circle to become gapless, i.e. the node must pass through the circle. Of course the above statement applies to {\it every} loop  enclosing the node. This leads to the ``vorticity conservation law'' - a {\em single} node can be displaced but not annihilated. Annihilation can occur only when two opposite vorticity nodes hit each other.
Therefore although ``accidental'' these nodes are stable against small perturbations.

\begin{figure}[tbp]
\begin{center}
\includegraphics[scale=0.5]{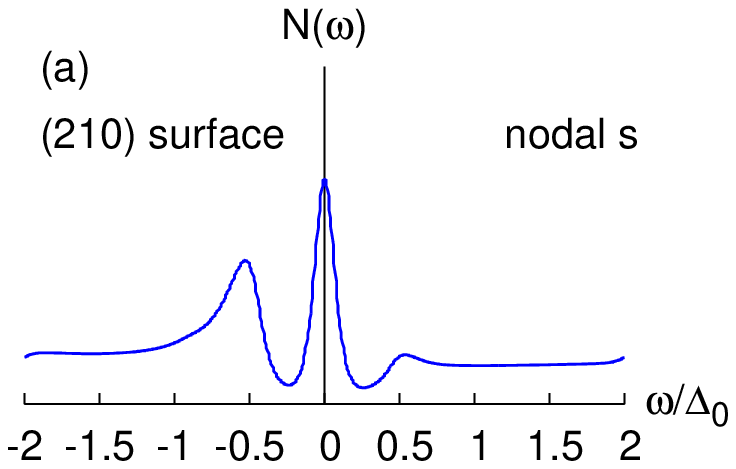}\includegraphics[scale=0.5]{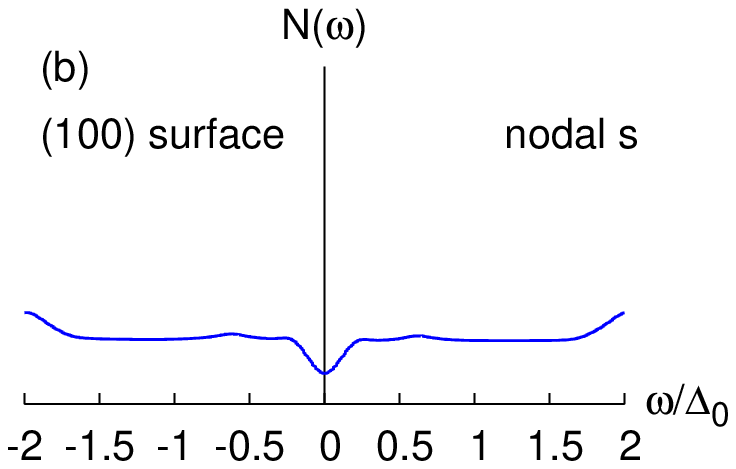}
\includegraphics[scale=0.5]{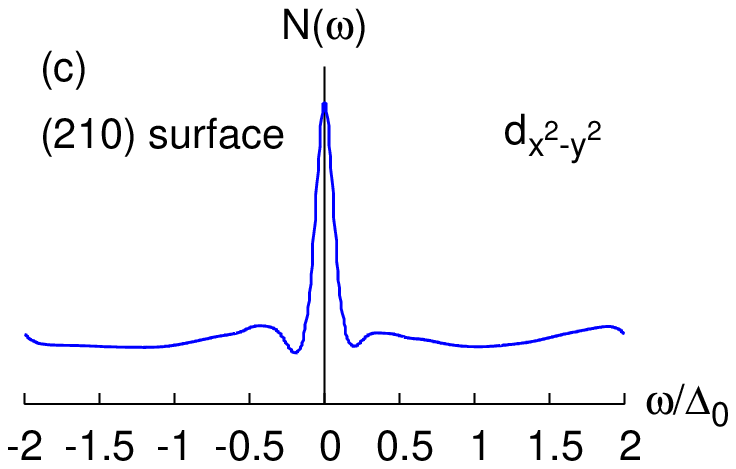}\includegraphics[scale=0.5]{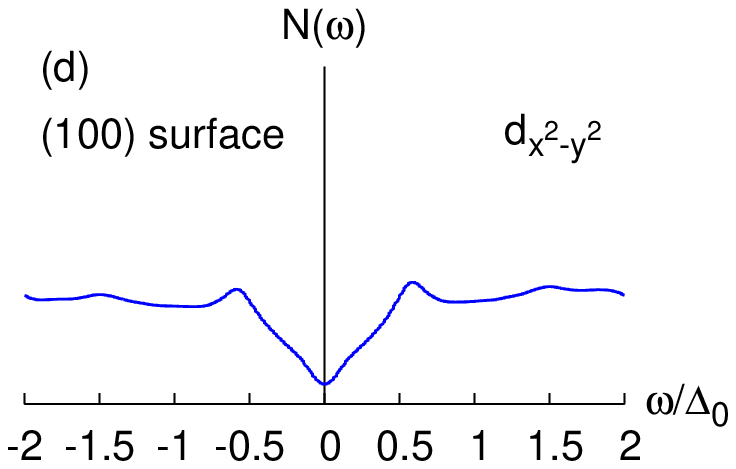}
\includegraphics[scale=0.5]{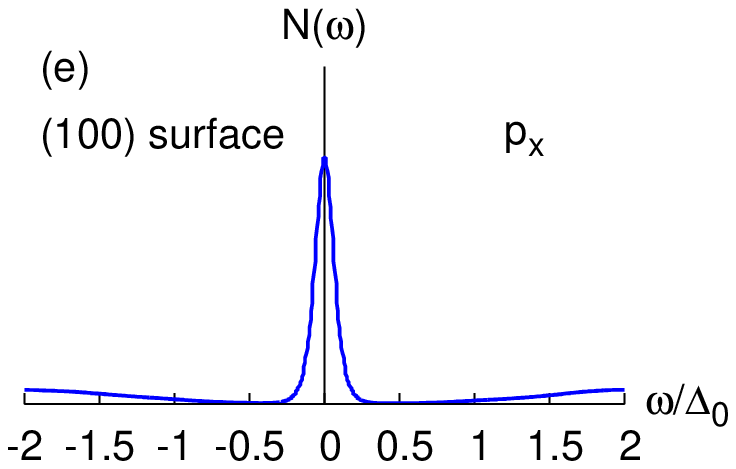}\includegraphics[scale=0.5]{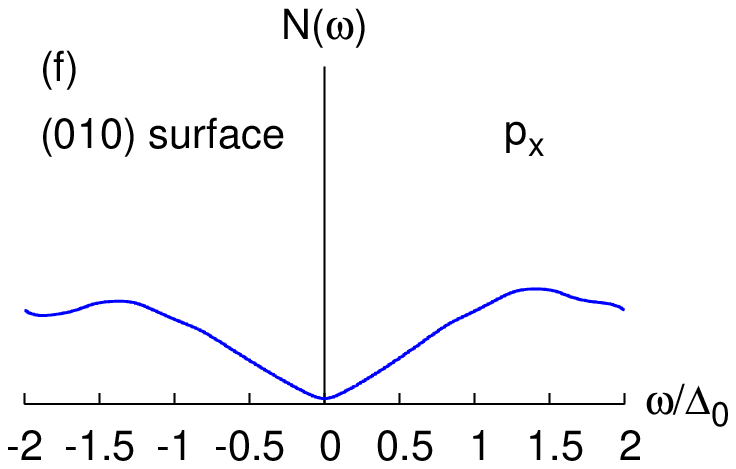}
\caption{
The surface local density of states $N(\omega)$ for some of the pairing symmetries in Table~\ref{t1}.
For each symmetry two surface orientations are shown: one with zero bias peak and the other without.
The unit of the density of states is arbitrary.
The $\Delta_0$ used in this calculation is $0.01$eV.
Thermal broadening with temperature $T=0.4$meV is used.
}
\label{fig2}
\end{center}
\end{figure}

Because the opposite vorticity nodes distribute symmetrically around the $(10)$, $(01)$ and $(11)$, $(1\bar{1})$ axes in \Fig{fig1}a, there will be no ZBABS on the edges. This is in sharp contrast to the d$_{x^2-y^2}$ nodes (\Fig{fig1}c) which possess ZBABS on the $(11)$ and $(1\bar{1})$ edges, as well as the d$_{xy}$ nodes which have ZBABS on the $(10)$ and $(01)$ edges. The lowest index edge that does show ZBABS are the $\{21\}$ edges. (The corresponding orientation of the real surfaces can be obtained by adding an extra zero, e.g., $(11)\ra (110)$ etc.)
Using the tight-binding fit of LDA bandstructure adapted from Ref.~\cite{acc1,acc2} and the following BdG Hamiltonian
\be
H_{\rm LaFePO}(\v k)=\epsilon(\v k)\otimes\s_0\otimes\tau_3+\Delta(\v k)I_{5\times5}\otimes\s_0\otimes\tau_1,\label{lfp}
\ee
we calculated the surface local density of states for a number of different {\em singlet} pairing symmetries and surface orientations. In \Eq{lfp}  $\epsilon(\v k)$ is the $5\times 5$ bandstructure matrix,
$\Delta(\v k)=2\Delta_0(\cos k_x+\cos k_y)$ and  $I_{5\times 5}$ is
the identity matrix in the orbital space.
(The calculation uses the iterative Green's function method of Ref.~\cite{algorithm}.)
A few examples of the results for nodal singlet gap function are shown in \Fig{fig2}a-d.

Of course the $s$-wave accidental nodes can fall on the central hole pockets too, an example is shown in \Fig{fig1}b. (Not dictated by symmetry, there is no reason for accidental nodes to occur on both hole Fermi surfaces.)
Due to the reflection symmetry about the $\{10\}$ and $\{11\}$ axes, the projection of the opposite vorticity nodes coincide, hence there will be no ZBABS on those surfaces. In contrast there is no reflection symmetry about the $\{21\}$ directions hence generically these edges will have ZBABS . This argument holds when the accidental nodes exist on both hole pockets and/or on the electron pockets.

Because the unusual temperature dependence of the Knight shift we have also considered triplet pairing. The BdG Hamiltonian for this case is given by
\be
H^\prime_{\rm LaFePO}(\v k)=\epsilon(\v k)\otimes\sigma_0\otimes\tau_3+I_{5\times 5}\otimes\Delta(\v k)\otimes\tau_1,
\label{equ:p-wave}
\ee
 where $\Delta(\v k)=\vec{d}(\v k)\cdot\vec{\s}$ with $\vec{d}(\v k)=-\vec{d}(-\v k)$.
In general if there is no further restriction on $\vec{d}(\v k)$ the node should not be topologically stable. This is because for a loop
consists of {\rm generic} momenta around a node, the mapping from $\v k$ to $(\epsilon(\v k),\vec{d}(\v k))$ ($\e(\v k)^2+|\vec{d}(\v k)|^2\ne 0$) has the homotopy group $\pi_1(S^3)=0$. The situation can change if we impose further constraints on the form of $\vec{d}(\v k)$. For example, ignoring the spin-orbit interaction, if we require  $\vec{d}(\v k)=f(\v k)\hat{d}_0$ where $f(\v k)$ is a partner function of the point group irreducible representation, the node becomes stable, and is characterized by the winding number of the AIII class\cite{ryu}. Under such condition we expect boundary zero energy states. For simplicity we considered only the following $p$-wave pairings with
$\vec{d}(\v k)=2\Delta_0\sin k_x~\hat{z}$ ($p_x$ pairing), $2\Delta_0\sin k_y~\hat{z}$ ($p_y$ pairing) and
$2\Delta_0\sin(k_x\pm k_y)\hat{z}$ ($p_{x\pm y}$ pairing).

In Table~\ref{t1} we summarize the results for eight different time reversal symmetric nodal pairing scenarios. In addition to the four triplet gap functions discussed above we have also studied
nodal $s$, $\Delta(\v k)=2\Delta_0(\cos k_x + \cos k_y)\s_0$;
$d_{x^2-y^2}$, $\Delta(\v k)=2\Delta_0(\cos k_x - \cos k_y)\s_0$ and
$d_{xy}$, $\Delta(\v k)=4\Delta_0 \sin k_x \sin k_y\s_0$ (here $\Delta_0=0.01$eV). Moreover  we have also considered the horizontal ring nodes observed in Ref.~\cite{feng} for BaFe$_2$(As$_{1-x}$P$_{x}$)$_2$. Surface orientation wise we considered
$(010)$, $(100)$, $(1\bar{1}0)$, $(110)$ and $(210)$.
Since there are no identical rows, by studying the dependence of the zero-bias peak on the surface orientation one can differentiate different types of nodal pairing symmetries.

\begin{table}
\caption{``$\checkmark$'' denotes the presence of ZBABS and ``$\times$'' the absence of it.
The pairing symmetries are defined in main text.
}
\begin{tabular}{|c|c|c|c|c|c|}
  \hline
  symmetry\textbackslash surface & $(010)$ & $(100)$ & $(1\bar{1}0)$ & $(110)$ & $(210)$ \\ \hline
 nodal $s$ & $\times$ & $\times$ & $\times$ &$\times$ &$\checkmark$ \\
  $d_{x^2-y^2}$ & $\times$ & $\times$ & $\checkmark$ & $\checkmark$ &  $\checkmark$ \\
  $d_{xy}$ & $\checkmark$ & $\checkmark$ & $\times$ & $\times$ & $\checkmark$ \\
    horizontal ring nodes & $\times$ & $\times$ & $\times$ & $\times$ & $\times$ \\
  $p_x$ & $\times$  &$\checkmark$ & $\checkmark$ & $\checkmark$ &$\checkmark$ \\
  $p_y$ & $\checkmark$ & $\times$  &$\checkmark$ & $\checkmark$ & $\checkmark$\\
  $p_{x+y}$ & $\checkmark$ & $\checkmark$ & $\times$ & $\checkmark$ &$\checkmark$ \\
  $p_{x-y}$ & $\checkmark$ & $\checkmark$ & $\checkmark$ & $\times$ & $\checkmark$\\
    \hline
 \end{tabular}\label{t1}
\end{table}

Although our discussion is in the context of LaFePO, 
our results should apply to other iron-based superconductors with the same
lattice symmetry and time-reversal symmetry as well. 
For example the proposed $d_{x^2-y^2}$-wave pairing in KFe$_2$As$_2$\cite{Thomale-KFeAs} 
should have the same surface orientation dependence of ZBABS as the cuprates.

{\bf{Effects of disorder.}}
A natural question one might ask in reading this work is how protected these E=0 surface states really are. Because $\tau_2$, the second Pauli matrix in the Nambu space, anticommutes with all Hamiltonians we have discussed, the E=0 states can be made eigenstates of it. It can be shown that for a fixed winding number the E=0 surface states with different $\tau_2$ eigenvalues localize at the opposite ends of the sample. Reversing the winding number exchanges these localized states. Thus local disorder potential can only mix E=0 states at surface momenta
where {\it opposite} winding numbers occur.  It turns out time reversal symmetry requires the 1D winding number to change sign from surface momentum $k$ to $-k$. As the result there are equal, opposite, intervals in the boundary BZ with opposite winding numbers. Smooth surface potentials do not mix them but sharp impurity potentials do. In the presence of the mixing the delta function density of states broadens into a peak centered around $E=0$. The width of the peak is proportional to the disorder strength. Therefore although the singular density of states at E=0 is not protected, a broadened version of it is. Interestingly the $p_{x,y}$-wave pairing studied in conjunction with LaFePO evades disorder broadening hence is strictly protected. This is because opposite winding number occurs for opposite spin. As long as the disorder potential does not flip spins the $E=0$ singularity is unaffected. Of course the ability to see the zero bias conductance peak\cite{zbcp1,zbcp2} and its surface orientation dependence\cite{ori1,ori2,ori3,ori4} in real experiment on cuprates, where disorder should mix ZBABS, attests for the robustness of the phenomenon discussed here.

\begin{figure}[tbp]
\begin{center}
\includegraphics[scale=.8]{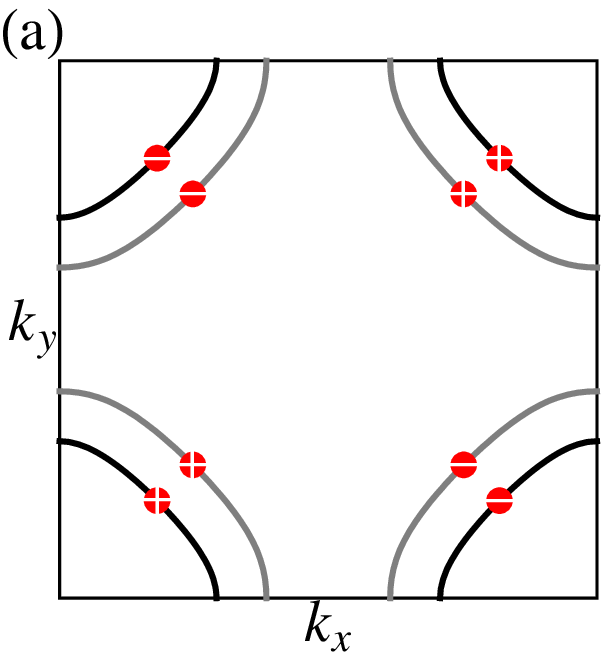}\hspace{0.1in}
\includegraphics[scale=.8]{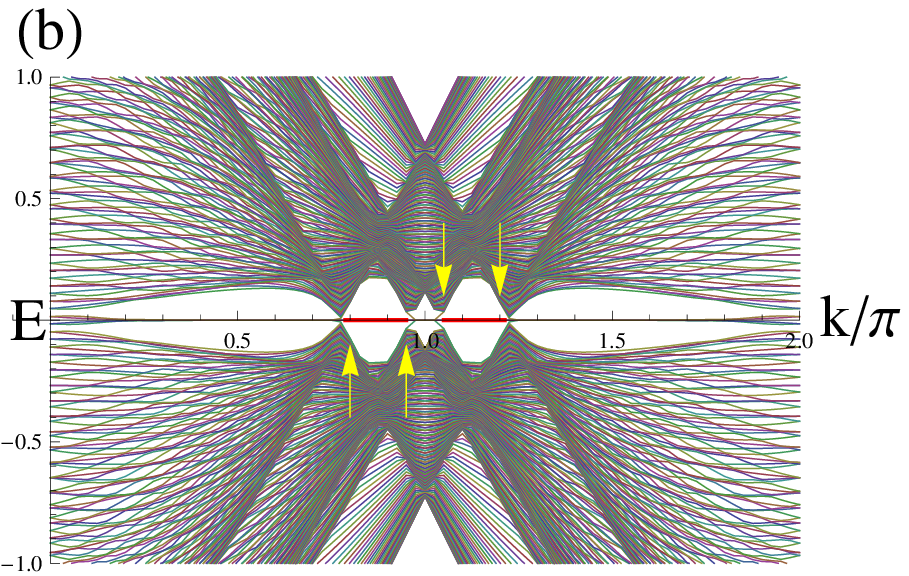}
\caption{(a) The spin non-degenerate Fermi surfaces (the black and gray curves) of the non-centrosymmetric superconductor defined in \Eq{nch}. The nodes are designated by the red dots. (b) The bandstructure of the $\{11\}$ edges. The
edge states exist between the momenta marked by the yellow arrows. The parameters we used in producing \Fig{fig3}b are $\lambda=0.3,\mu=0.45$, $\Delta_0=0.1$ and $\alpha=1,\beta=0$.}
\label{fig3}
\end{center}
\end{figure}

{\bf{Nodal non-centrosymmetric superconductors.}}
Another interesting application of the idea presented here is to nodal non-centrosymmetric superconductors\cite{noncentro}. In these systems due to the spin-orbit interaction the Fermi surfaces are spin non-degenerate. This is shown by the black and gray curves in \Fig{fig3}a. Several known non-centrosymmetric superconductors exist near an antiferromagnetic phase. These include CePt$_3$Si, CeRhSi$_3$, CeIrSi$_3$ and CeCoGe$_3$\cite{noncentro}. For these systems it is natural to expect the superconducting gap function to possess nodes (in particular d-wave nodes). This is indeed substantiated by experimental findings\cite{ncnode}. Therefore according to earlier arguments there should be surface E=0 bound states. We consider the following simple model BdG Hamiltonian
\be
\begin{split}
H(\v k)=
\,&
\epsilon(\v k)\s_0\otimes\tau_3-\lambda\vec{\gamma}(\v k)\cdot\vec{\s}\otimes\tau_3
\\&
+\Delta_0
(\cos k_x-\cos k_y)[\alpha \s_0+\beta \vec{\gamma}(\v k)\cdot\vec{\s}]\otimes\tau_1.
\end{split}
\label{nch}
\ee
Here $\epsilon(\v k)=-\cos k_x-\cos k_y-\mu$,
$\vec{\gamma}(\v k)=(\sin k_y,-\sin k_x)$ and $\vec{\s}=(\s_1,\s_2)$.
The bandstructure on the $\{11\}$ edges of this model Hamiltonian is shown in \Fig{fig3}b. There are E=0 flat bands in different portions of the edge BZ. Generally, when exist, such flat bands are two fold degenerate, i.e., on each edge there are two $E=0$ bound states. The only exception are the flat bands (marked by the red thick line segments in \Fig{fig3}b) in the momentum intervals between the yellow arrows. The momenta of these two intervals are opposite to each other (up to reciprocal lattice vector). In addition the zero energy bound states there are {\em non-degenerate}. The E=0 quasiparticle operator associated with the two different red intervals are hermitian conjugate of each other.
In Ref.~\cite{ryu2} similar surface flat bands are found in the nodal states bordering non-centrosymmetric (fully gapped) topological superconductors.\\

{\bf{Detecting nodal spin liquids.}}
The search for spin liquid is a focused interest in current condensed matter physics. The difficulty with finding a spin liquid is precisely why they are novel - they are ``featureless'' in conventional sense. Among different spin liquids there is a class that are nodal (to ensure their stability we focus on the ones having $Z_2$ gauge symmetry).  For example such a spin liquid was proposed for the pseudogap state of the cuprates\cite{Senthil-Fisher, WenXG-Z2}. Like nodal superconductors described above, nodal spin liquid should have surface ``spinon'' flat bands.
If exist these flat bands will give rise to Curie behavior of local magnetic susceptibility hence allows for experimental detection, for example the low temperature susceptibility will diverge as $1/T$.
Of course, as discussed earlier, disorder broadens the singular density of state at $E=0$ into a peak hence renders the local magnetic susceptibility finite. In addition one might need to worry about the effect of residual spinon interaction on these flat bands. We believe if the residual spinon interaction is repulsive surface ferromagnetism will result.

{\bf{Discussions.}}
Experimentally one can prepare differently oriented surfaces by thin film growth and perform scanning tunneling microscopy (STM) measurements to measure the local density of states associated with ZBABS. The advantage of using STM is it can probe $\sim nm$  length scale.  Therefore even when the surface is not ideal, it can pick up signals from surfaces with different local orientations.
Thus by scanning the tip it is possible to acquire the required data from a single sample.

After completion of this work we noticed a very recent experiment observing zero-bias conductance peak for FeSe$_{0.3}$Te$_{0.7}$\cite{WuMK}.

\acknowledgments{
We are very grateful to Volovik for pointing out his recent work\cite{Volovik}
where similar argument was presented.
We thank Yuan-Ming Lu, Kyle Shen, and Nai-Chang Yeh for very helpful discussions.
DHL acknowledges the support by the DOE grant
number DE-AC02-05CH11231.}

\begin{appendix}
\section{The criterion illustrated by the cuprate example}
\label{app:example}

\begin{figure}
\begin{center}
\includegraphics[scale=.7]{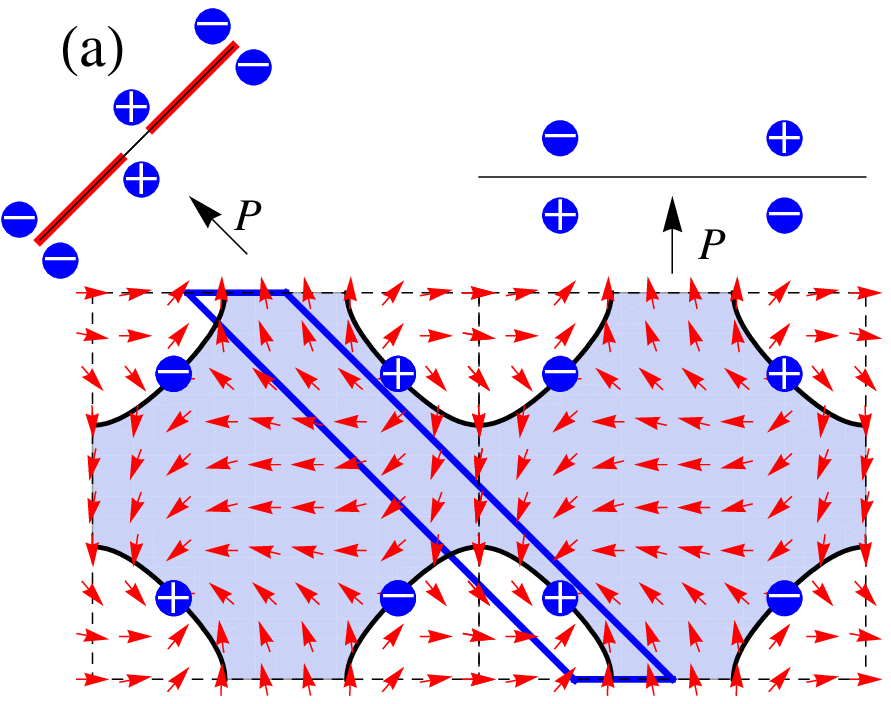}
\includegraphics[scale=.7]{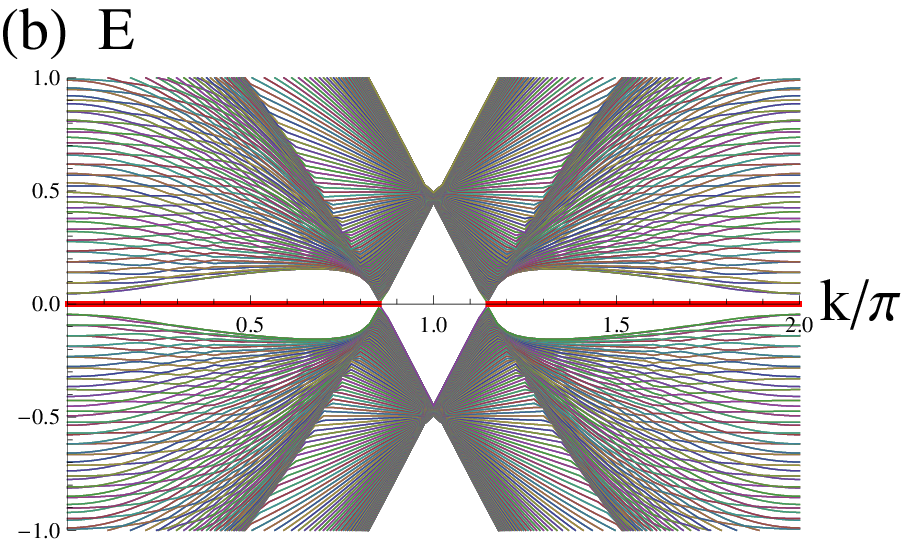}
\caption{(a)~The 2D nodal band structure and its projections on $(1\bar{1})$ and $(01)$
surfaces for a model $d$-wave superconductor. The first BZ and one extended zone are drawn. The solid black curves denote the Fermi surface and the shaded region is filled in the normal state.
Red arrows are the unit vectors in \Eq{n}. The blue dots with $\pm$ signs represent nodes with
vorticity $\pm 2$ respectively. Top left: the slanted thin black line segment is the surface
BZ of the $(1\bar{1})$ edge. The black arrow with letter ``P'' indicates the direction of projection.
Thick red line segments mark the surface momenta with zero energy bound states. Top right:
the thin black horizontal line segment is the BZ of the $(01)$ edge.
The $\pm$ vorticity nodes overlap after projection.
The  vorticity of the node enclosed by the parallelogram is equal to the winding number difference along the two side blue segments because the winding along the top and bottom segments cancel due to the periodicity in momentum space. This can be explicitly seen by following the turning of the red arrows (the actual winding number is twice of the winding shown by the arrows, due to the spin degeneracy). (b) The (11) edge bandstructure. The surface flat bands are marked red.  In constructing this figure we have used $\epsilon(\v k)=-\cos k_x-\cos k_y-\mu$ and $\Delta(\v k)=\Delta_0 (\cos k_x-\cos k_y)$ in \Eq{dwave}. Here  $\mu=0.45,\Delta_0=0.1$.}
\label{fig:d-wave}
\end{center}
\end{figure}

The idea behind the criterion presented in the main text is 
best illustrated by using the cuprate superconductor as an example.
The Bogoliubov-de Gennes (BdG) Hamiltonian of the cuprate superconductor read
\be
H_{\rm cuprates}(\v k)=\epsilon(\v k)\s_0\otimes\tau_3+\Delta(\v k)\s_0\otimes\tau_1,\label{dwave}
\ee
where $\s_0$ is the identity matrix acting in the spin space and $\tau_{1,3}$ are $2\times 2$ Pauli matrices in the Nambu space, $\epsilon(\v k)$ is the normal state dispersion satisfying $\e(-\v k)=\e(\v k)$ and $\Delta(\v k)$ is the d-wave gap function. (Since the cuprates are quasi two dimensional materials, we shall use two dimensional notations in the following discussions.) The Fermi surface and the gap nodes are shown in \Fig{fig:d-wave}a, therefore $d=2,q=0$. In the same figure, the normalized vector  \be \hat{n}(\v k)=(\epsilon(\v k),\Delta(\v k))/\sqrt{\epsilon(\v k)^2+\Delta(\v k)^2}\label{n}\ee is plotted as a function of $\v k$ over two BZs (see red arrows). Inspecting these arrows one notices each node is a ``vortex'' in $\hat{n}(\v k)$. Around each vortex the arrows exhibit non-zero winding. The total winding number associated with each node is given by 
\be
w={2\over 2\pi}\oint d\v p\cdot [n_1(\v k)\gr_{\v k} n_2(\v k)-n_2(\v k)\gr_{\v k} n_1(\v k)].
\label{w}
\ee (The extra factor of 2 is due to spin degeneracy). Clearly each node is characterized by an even integer winding number. The BdG Hamiltonian defined on all one ($=d-q-1$) dimensional loop enclosing the node are topologically nontrivial.

Now consider the bandstructure projected along the $(1\bar{1})$ direction. For each transverse momentum $k$ along $(11)$ we have a 1D chain running in $(1\bar{1})$.
So long as $k$ does not coincide with the projection of the nodes the spectrum is fully gapped and characterized by the winding number defined in \Eq{w}. Any two chains whose $k$ straddle the projection of a node their winding numbers must differ by $\pm 2$ (see \Fig{fig:d-wave}a captions), hence at least one of them is topologically non-trivial and possess $E=0$ end states
when the boundary condition along $(1\bar{1})$ changes from closed to open.
~This implies  E=0 bound states exists for {\em intervals} of $k$. Therefore $d_{E=0}$ is indeed $q+1=1$. An example of the (11) boundary bandstructure is shown in \Fig{fig:d-wave}b. The $k$ intervals showing the flat bands are represented by the thick red line segments in the top left corner of \Fig{fig:d-wave}a.  The only edges which do not possess ZBABS are the $\{10\}$ (Miller's notation is used) edges where the projection of positive and negative nodes overlap (see top right corner of \Fig{fig:d-wave}a).
For the real material $d=3, q=1$ and the only modification is $d_{E=0}$ changes from 1 to 2.

\end{appendix}

\end{document}